# Analysis of Current Hysteresis in Graphene Field Effect Transistor

Yifan Yao[1]

Department of Material Science and Engineering, University of Illinois at Urbana-Champaign,

Urbana 61801, United States

## Abstract

**In this report, the hysteresis behaviors of Poly-ethylenimine(PEI): LiClO4 and Poly-ethylene-glycol(PEG): LiClO4 electrolyte gate and back gate Graphene-on-SiO2 FET (GFET) were analyzed by gate voltage—source-drain current modulation. It is shown that both the sweeping rate and the sweep range will cause hysteresis behaviors in the form of Dirac Point ($V_{dp}$) shifting or changes in the current. Different mechanisms including charge trapping and electrical double layer capacitive effect are proposed to explain the behavior qualitatively on both back gated and electrolyte gated FET and partially confirmed with the present experimental results.**

[1] Email: yyao13@illinois.edu; Phone: 2179792996

# 1. Introduction

Graphene has been attracting a great amount of interest since its discovery and nowadays more efforts are made to utilize its unique electronic and mechanical properties, including the tunable band gap, high electron mobility, great mechanical robustness by integrating graphene-based transistors. [1]. The density of states (DOS) of graphene is dependent on the fermi level, which can be modulated by the applied voltage and theoretically reaches zero at the Dirac point at which no gate-induced carriers are present. In the normal operation mode, when the gate voltage is equal to the Dirac point, the conductance will reach minimum under ideal conditions [2]. However, a hysteresis effect will cause a shift in the Dirac point position and the conductance under forward gate voltage sweep and backward sweep. [3] This issue is of the great importance before we can fully understand the behavior of GFET although there is no universal agreement on the cause of such effect.

Several mechanisms have been proposed to explain its origin including charge trapping related to the surrounding environment. [4]. In the case of electrolyte gate GFET, an ionic double layer is formed to be the capacitor when applying a gate voltage [5]. The hysteresis behavior of the electrolyte gate GFET was therefore believed to be relevant to the constraint movements of ions in the polymer solution. [3]. In the present work, the current hysteresis of electrolyte gated transistor and back gated transistors were investigated with these mechanisms and compared to the experimental results of Graphene-on-SiO2 back gated FET with intrinsic/doped Graphene respectively and LiClO4 electrolyte gated FET in two different polymer solutions: PEI and PEG. New experiments are proposed to look further into the physical origin and expected to have more direct proof of the hysteresis hypothesis.

# 2. Experimental Methods

**2.1 Graphene Layer Wet Transfer:** The graphene channel in Back Gated GFET was transferred from commercially available Cu/CVD Graphene/PMMA pieces. The Cu coil layer was etched by 1M ammonium persulfate solution for 15 minutes and Graphene/PMMA layer was bathed for 3 times in DI water, each with 5 minutes. The Graphene/PMMA was then transferred onto the pre-sonicated Si/SiO2 substrate and dried at room temperature. The PMMA layer was removed by dissolving in $1^{st}$ acetone solution for 10 minutes and $2^{nd}$ acetone solution for 1 hour. Graphene layer was retrieved on Si/SiO2 after degreasing and cleaning with PMMA removed.

**2.2 Back Gated GFET Fabrication:** Back Gated GFET was fabricated by depositing Ti-Au Electrode on the Graphene-on-SiO2/Si sample made in Method 1 by sputtering.

**2.3 Polymer Electrolyte Solution Preparation:** Both PEI: LiClO4 and PEG: LiClO4 were prepared at 3:1 weight ratio and sonicated for 5 minutes.

**2.4 Electrolyte Gated GFET Fabrication:** Electrolyte Gated GFET was fabricated using the commercial Si/SiO2/Graphene samples. The source and drain contact was made by contacting indium wires (1'' long) + electrical wires (3" long) onto 1mm droplet of 1:1 AB silver epoxy. Epoxied wires were cured at 60°C for 30 minutes.

**2.5 Electrical Characterization of Back Gated GFET:** Back Gated FET was electrically characterized on a common two terminal probe station. Details in operating the station are not described here. After the characterization of the back gated device, 1mm droplet of commercial PEI was doped on the graphene channel. The doped device was measured after 10 minutes.

**2.6 Electrical Characterization of Electrolyte Gated GFET:** Similar characterization was carried out on the electrolyte gated FET on an SMU. A droplet of (~1mm) PEG: LiClO4 from Method 3 was dropped between electrodes and a Pt wire was connected to the gate to carry out the measurement. PEG: LiClO4 was washed away with methanol, acetone and isopropanol in sequence. A measurement was carried out after dropping 1mm PEI: LiClO4 from Method 3 similarly.

## 3. Results & Discussion

In this report, the hysteresis behavior of Graphene field effect transistor (GFET) was analyzed by sweeping from a negative gate voltage to a positive gate voltage (a forward scan), immediately followed by a backward sweep. The mechanism of hysteresis was explored by varying the sweep range and the sweep rate and studied individually.

## 3.1 Graphene-on-SiO2 back gated GFET

The hysteresis behavior of the SiO2 back gated graphene transistor is shown in Figure 1. To characterize the hysteresis quantitatively under different sweep rates, either the loop area enclosed by the transfer curve or the change in current when the gate voltage is zero can be used for comparison since all voltage-current curves share similar shapes. From Figure 1(b), the loop area and the change in current matches well with each other, a good indicator that transfer curve shapes are similar for different sweep rates. From Figure 1(c) and 1(d) it can be seen when the sweep range increases from 20V to 80V, the hysteresis increases as well, which can be characterized by the change in current when the gate voltage is zero since all forward currents under are at the similar magnitude. The positive correlation between the sweeping range and the hysteresis can be explained by the charge trapping mechanism [3]. Due to the defects pre-existed in the graphene film, there are defect sites that trap electrons and holes as their densities are modulated by the back-gate voltage. When the voltage sweeps in the positive regime, electrons induced in the graphene will likely to be trapped at so-called 'charge traps' and when the voltage decreases, so the graphene will become more negative than it should be due to the only gate voltage modulation, theoretically pushing the charge neutrality point (CNP) in the more positive voltage direction under such sweep. The polarity of the charge trap will be reversed when the voltage sweeps into the negative regime. A simple model can be used to estimate the number of charges trapped:

$$n = \frac{\Delta V_{dp} C}{2e} \qquad\qquad Equ.1$$

which $\Delta V_{dp}$ is the change in Dirac Point and C is the gate capacitance (calculations shown in Supporting Information). Both the hysteresis "direction" (CNP shifting direction) and magnitude (current different between two sweeps under different sweep ranges) can be considered in this case. When the CNP is not shown in Figure 1 (possibly larger than 40V due to ambient doping) [6, 5], the CNP shift is theoretically comparable to the upward shift of the current in the reverse sweep (from positive to negative voltage) when Vg is zero, which is agreed with the present experimental result. The linear relationship between the hysteresis and the sweep range might be explained by the hypothesis that the number of trap charges is relatively proportional to the electrical field applied to the graphene channel.

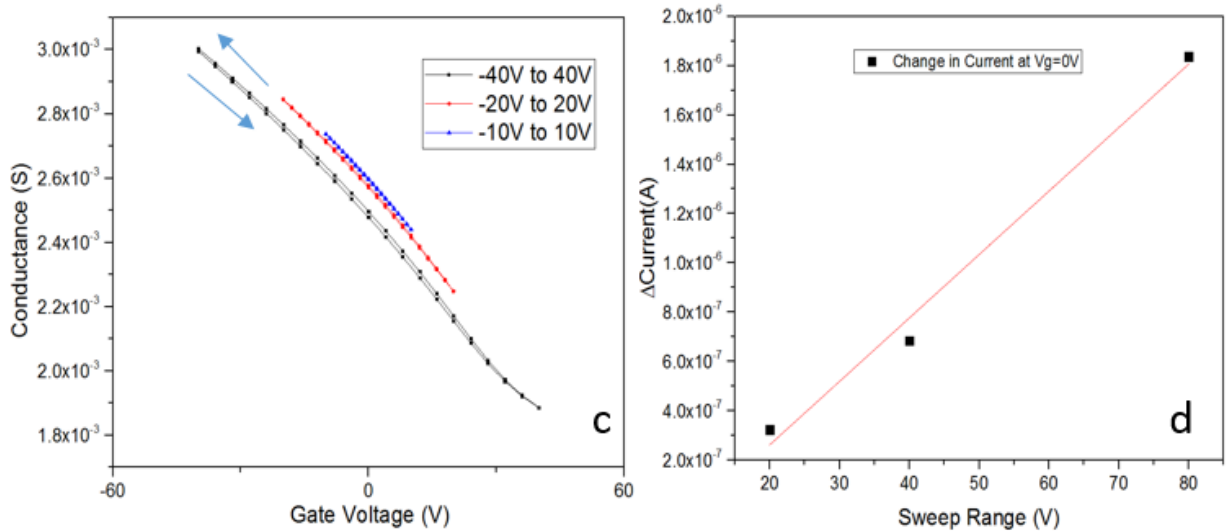

*Figure 1 (a): Conductance change vs. gate voltage at different sweeping rates from -40V to 40V of Graphene-on-SiO2 back gated transistor (b): Difference in current without the back-gate voltage between the forward and the backward sweep and the 'loop area' enclosed the conductance curve at each sweeping rate from (a). (c): Conductance change vs. gate voltage under different sweeping range at 3.77V/s sweep rate. (d): Change in current from (c) between backward and forward scan at no gate voltage.*

From Figure 1(a) and 1(b), there is very limited hysteresis at each sweeping rate from 3.77V/s to 109.58V/s in the sweep range of -40V to 40V with almost no obvious relationship between the sweeping rate and the hysteresis. However, it can also be the result of two competing mechanisms: ambient doping and charge trapping. The effect of charge trapping is more obvious when graphene is doped by Poly-ethylenimine(PEI). In Figure 2(C) and 1(C), there is a negative correlation between the hysteresis created by charge trapping and the sweep rate. This relationship is also observed in other similar experiments and is attributed to the long time scale of charge trapping. [3]. The conductance behavior at the sweep rate higher than 100V/s is deviated from the inverse power relationship for un-doped graphene back gated device which can be explained by that when the hysteresis originated from the charge trapping is becoming less prominent as the sweep rate goes higher, other origins of the hysteresis will take over such as ambient doping effect. [7]. It has also been shown the hysteresis from ambient doping can be effectively eliminated when a polymer electrolyte such as PEI and PEO are put on the graphene channel. [8], which is confirmed by our results.

## 3.2 PEI-Graphene-on-SiO2 back gated GFET

Other similar results can also be observed when the same back gated device is doped by PEI. Such hysteresis behavior is shown in Figure 2. The linear relationship between the sweep range and the hysteresis and the inverse power relationship is also shown in Figure 2(b). The agreement between the loop area and the change in drain current at Vg =0 V still indicates the similar transfer curve shape in Figure 2(a) for different rates.

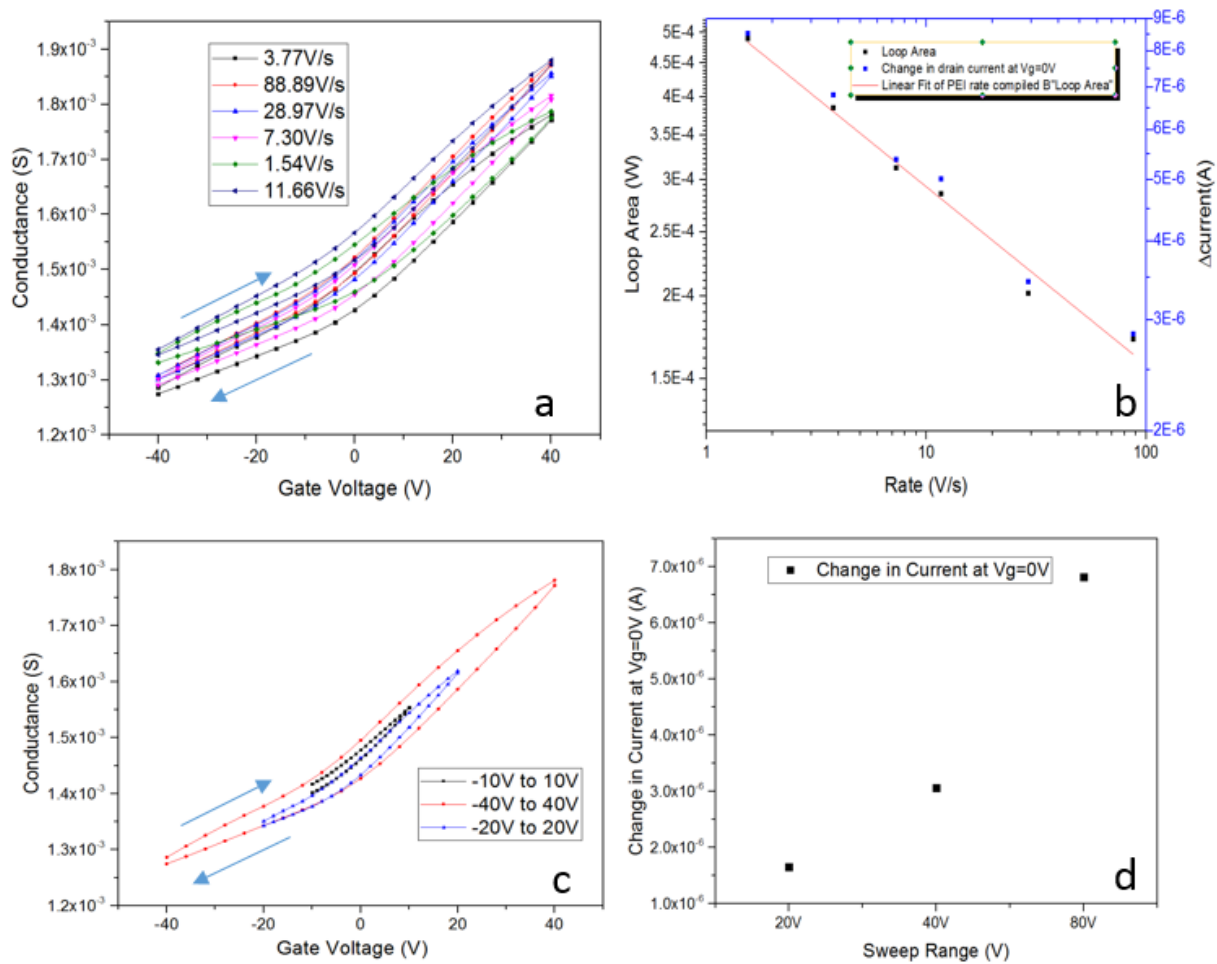

*Figure 2: (a) Conductance change vs. gate voltage at different sweeping rates from -40V to 40V of the same device used in Figure 1 doped with PEI (b) Difference in current without the back-gate voltage between forward and backward sweep and the 'loop area' enclosed the conductance curve at each sweeping rate from (a). (c) Conductance change vs. gate voltage under different sweeping range at 3.77V/s sweep rate. (d) Change in current from (c) between backward and forward scan at no gate voltage.*

## 3.3 PEI/PEG: LiClO4 electrolyte gated GFET

Other origins of the hysteresis behavior related to the polymer electrolyte gated graphene FET are also included. Shown in Figure 3 is the result of the PEI: LiClO4 electrolyte gated GFET, which was fabricated by Method 4. From Figure 3(a), a negative shift of DP is observed as the gate voltage sweeps from negative to positive and then to negative. At the sweeping rate of 0.99V/s, Vdp is equal to 0V under the forward sweep and shifts to -0.56V under backward sweep. This phenomenon can be explained qualitatively. After applying a positive voltage in the electrolyte solution after the first sweep, the dipole oriented polymer molecules and solvated ions will accumulate at the Graphene/electrolyte interface and form what is called an "Helmholtz double layer" capacitance, inducing carriers (electrons) in the graphene. [9]. The capacitance can be calculated using Equation 2:

$$C = \frac{\varepsilon_0 \varepsilon_r}{L_{db}} \qquad\qquad Equ.\,2$$

in which $\varepsilon_r$ is the dialectic constant of PEI and $L_{db}$ is the Debye Length. [5]. $L_{db}$ is dependent on the electrolyte concentration (ionic strength), which cannot be accurately determined when formed complexes with polymer chains. [10] It can be approximated to be 11.9* $10^{-6}$F/cm$^{-2}$ (calculation shown in Supporting Information) which agrees with the reported range as tens of $10^{-6}$F/cm$^{-2}$ [11] [12]. When sweeping backwards from a positive to a negative voltage, the Graphene/electrolyte interface will "remember" the ions accumulated in the first sweep and cause delay in the movement of these ions, thus creating a larger ion concentration at the interface in the reverse sweep than it should have at the same back gated voltage, which in turn will pushing the current upward in the positive regime and the Dirac point to the more negative regime. This "remembering" effect is associated with the limited ion mobility in PEI polymer matrix. Assuming the concentration gradient of ions at the interface is steep enough that can be modeled as a plated capacitor, the number of ions that are "remembered" can be approximated by Equation 1 in which $\Delta V_{dp}$ is the change in Dirac Point and C is the electrolyte gated capacitance (calculations shown in Supporting Information). In Figure 3(b), the difference in Dirac points between two sweeps is much larger as well as the enclosed area by the conductance curve with increasing sweep rates, indicating a positive correlation between the sweep rate and the hysteresis, which agrees with the capacitive gating hypothesis. It can be further deduced that the relaxation time of ions movement is comparable to the time scale of the testing sweep rates.

When tuning the gate voltage, we are tuning the Fermi Level ($V_f$) (the carrier density). When $V_f$ is equal to the charge neutrality point (Vdp), the minimum conductance is reached when there is an equal number of electrons and holes in the graphene underneath. The conductance can be approximated by Equation 3:

$$\sigma = n\mu_n e + p\mu_p e \qquad\qquad Equ.\,3$$

and is directly proportional to the total number of carriers. Comparing the charge neutrality point at different sweeping rates, the minimum conductance increases with increasing sweeping rates.

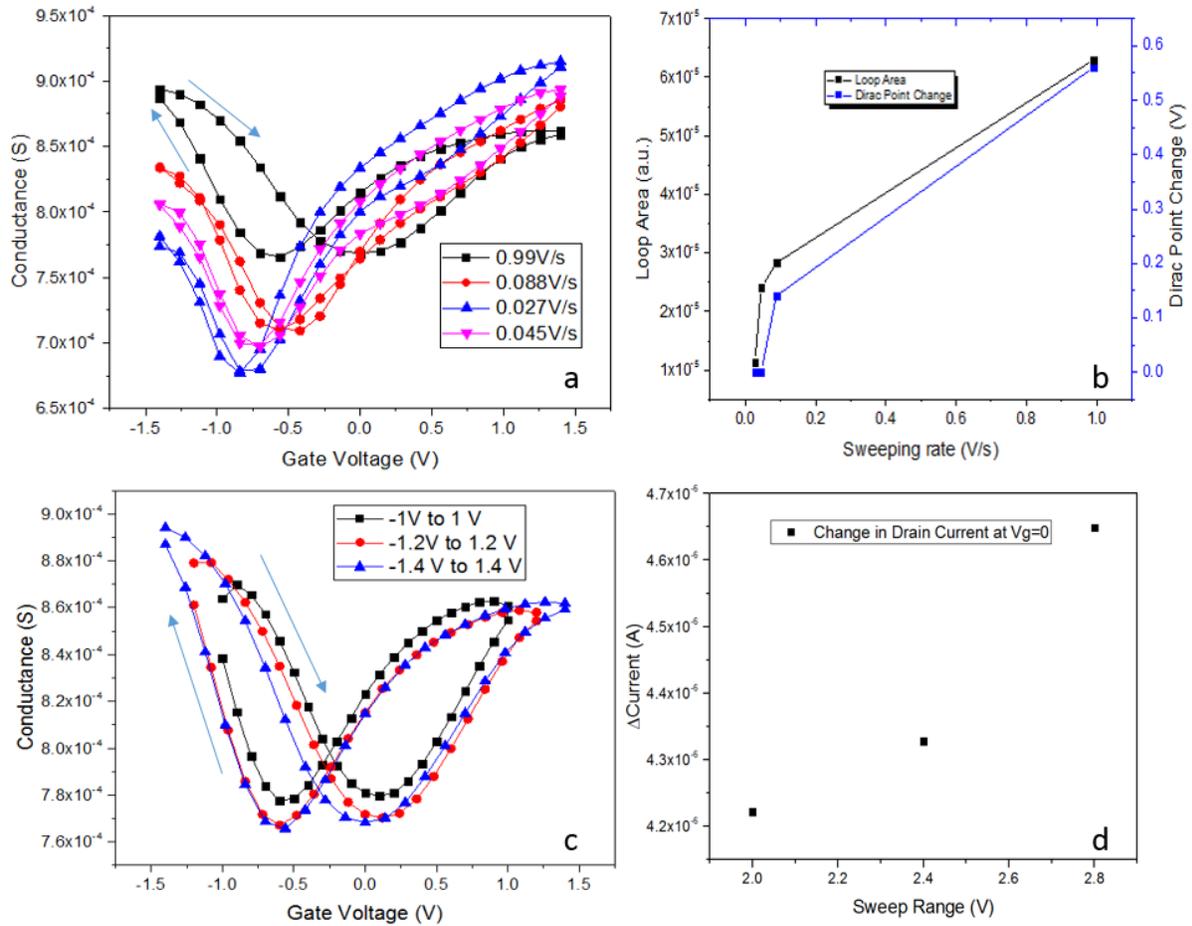

*Figure 3: (a): Conductance change vs. gate voltage at different sweeping rates from -1.4V to 1.4V of PEI: LiClO4 electrolyte gated GFET (b): Dirac Point shift between a backward and forward scan and the 'loop area' enclosed the conductance curve at each sweeping rate from (a). (c): Conductance change vs. gate voltage under different sweeping range at 0.99V/s sweep rate. (d): Change in current from (c) between backward and forward scan at no gate voltage.*

This indirectly supports of the "remembering" effect since under the same positive voltage but with elevated rate, result in more carriers and higher conduction at the elevated sweeping rate at the charge neutrality point. From Figure 3(c) and (d), the Dirac point seems to be fixing at 0.14V and -0.56V regardless of the sweep range. Each conductance curve seems to follow the same trend although the difference of the drain current at zero gate voltage is slightly increased with increasing sweeping range. Similar relationship is observed between the sweep range and the change in the current just like the one existed in back-gated devices, indicating that the charge trapping effect on the hysteresis might also be present in electrolyte gated device.

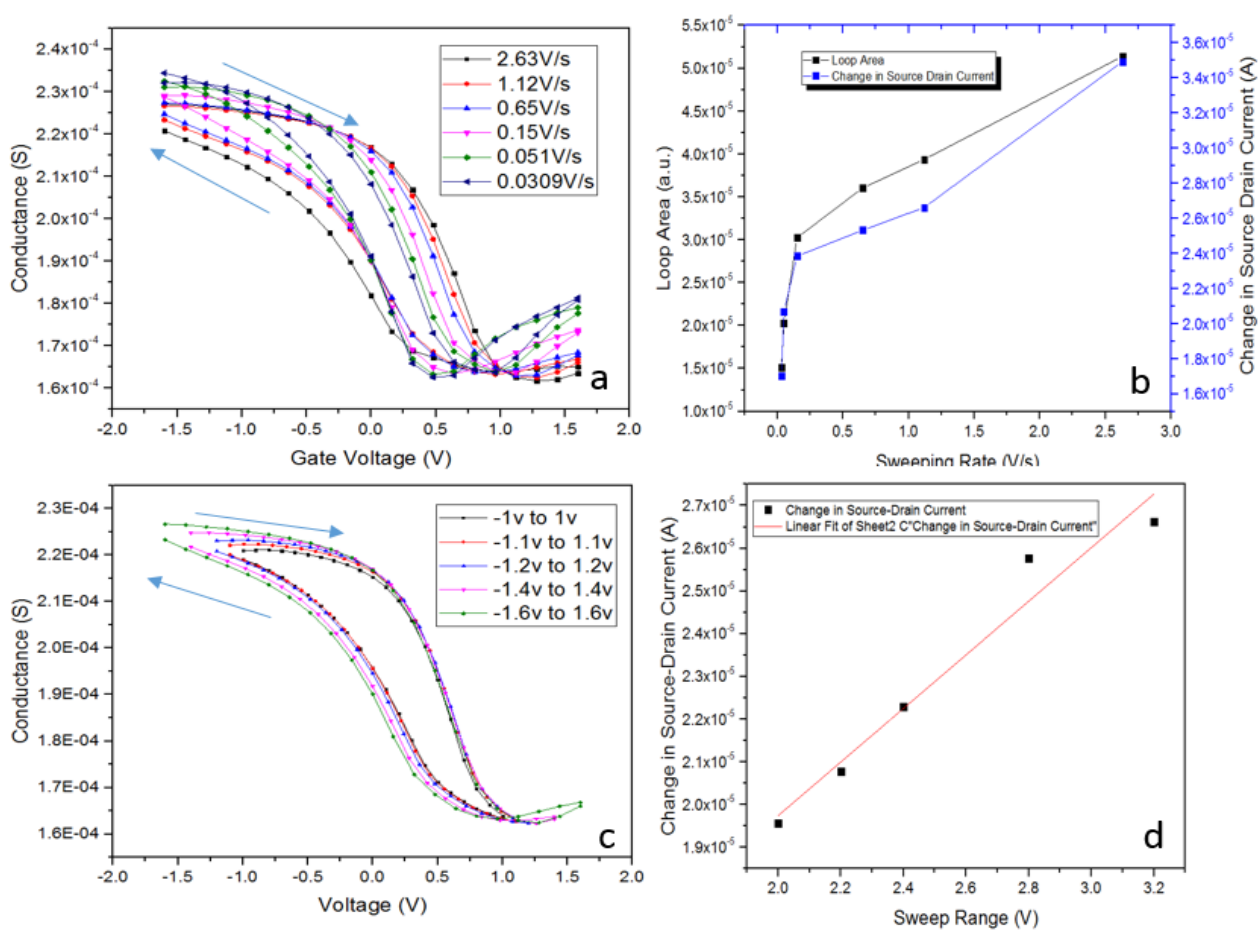

.

*Figure 4: (a): Conductance change vs. gate voltage at different sweeping rates from -1.6V to 1.6V of PEG: LiClO4 electrolyte gated GFET (b): Dirac Point shift between the backward and forward scan and the 'loop area' enclosed the conductance curve at each sweeping rate from (a). (c): Conductance change vs. gate voltage under different sweeping range at 0.99V/s sweep rate. (d): Change in current from (c) between backward and forward scan at no gate voltage.*

All the above effects existed in PEI: LiClO4 electrolyte gated device are also observed in Poly-ethylene-glycol(PEG): LiClO4 electrolyte gated device, as shown in Figure 4, including but not limited to: (1) a negative shift of DP between the forward sweep and the backward sweep (2) a positive correlation between the sweep rate and the hysteresis like in Figure3(b) (3). A similarly positive correlation between the sweep range and the hysteresis.

## 4. Conclusions

The Charge Trapping and the double layer capacitive gating are two mechanisms that can cause hysteresis and are examined with the present studies. The charge trapping effect is originated from trapping sites that trap mobile carriers while the electrical double layer capacitance effect results from the limited electrolytic ion movement in the matrix of large polymers. Both effects can be approximated by the equation: $n = \Delta V_{dp} C / 2e$., in which n is the trapped carriers in the charge trapping effect and the "delayed" ions in the capacitive gating effect. The charge trapping effect pushes the NP in the positive direction while the double layer capacitive gating effect pushes the NP in the negative direction under a forward and backward sweep, starting from a negative voltage to a positive voltage then to a negative voltage. Generally, faster sweep rates correspond to smaller hysteresis created by charge trapping and larger hysteresis created by the electrical double layer capacitance. A larger sweep range corresponds to a larger charge trapping as well. The hysteresis behavior of the back-gated device is dominated by the charge trapping mechanism (more prominent if the graphene is doped in PEI). However, the hysteresis behavior of the electrolyte gated device is the result of two competing mechanisms: both charge trapping and electrical double layer capacitance.

New experiments can be proposed on looking further into the mechanisms of two effects and providing more quantitative analysis. For the double layer capacitive effect, the electrolyte mobility in the polymer solution can be the focus of new experiments. The movement of ions and molecules in a polymer matrix is influenced by the concentration gradient (ionic strength) and mobility. In particular, ionic mobility in a polymer can be affected by temperature under the same sweep conditions by the Arrhenius equation: $\mu = \mu_0 \exp(-\frac{E}{RT})$ in which E is the arbitrary activation energy. The mobility could be correlated to the delay of the concentration change and measured by the electrolyte cell capacitance charging effect [13]. With appropriate modeling of the electrical double layer based on Gouy-Chapman-Stern analysis [14], the excess carrier density induced in graphene could also be approximated. We can thus build the relationship between the temperature change and the hysteresis based on the mobility change. The mobility also associates with the concentration (ionic strength) of the electrolyte ions because of the frictional force. A correlation can also be established between the ionic strength and the magnitude of hysteresis based on the mobility change of electrolyte ions. These proposed experiments are expected to provide a more direct way to prove the capacitive gating mechanism and a quantitative way to characterize its effect. [15]

## 6. Appendix

### 6.1 Gate Capacitance Calculation:

For Back gated GFET:

$$C = \frac{\varepsilon_0 \varepsilon_r}{t} = \frac{3.9 * 8.85 * 10^{-12} F/m}{285 nm} = 1.2 * 10^{-8} \frac{F}{cm^2} \qquad equ.\,4$$

For PEI:LiClO4 Electrolyte Gated GFET:

$$\frac{1}{C} = \frac{1}{C_q} + \frac{1}{C_{el}} \qquad equ.\,5$$

$$\frac{1}{C} = \frac{1}{2\,uFcm^{-2}} + \frac{L_{db}}{\varepsilon_0 \varepsilon_r} \qquad equ.\,6$$

PEI:LiClO4

$$L_{db} = \left(\frac{2ce^2}{\varepsilon\varepsilon0kT}\right)^{-\left(\frac{1}{2}\right)} = 13.33 nm \qquad equ.\,7$$

$$C = \left(\frac{1}{2\,uFcm^{-2}} + \frac{1}{0.21 cm^{-2}}\right)^{-1} = 1.19 * 10^{-5} \; F\,cm^{-2} \qquad eq\square.\,8$$

PEG:LiClO4

$$L_{db} = \left(\frac{2ce^2}{\varepsilon\varepsilon0kT}\right)^{-\left(\frac{1}{2}\right)} = 14.98 nm \qquad equ.\,9$$

$$C = \left(\frac{1}{2\,uFcm^{-2}} + \frac{1}{0.18\,uFcm^{-2}}\right)^{-1} = 1.15 * 10^{-5} F/\,cm^{\wedge}2 \qquad equ.\,10$$

## 6.2 Density of "remembered" ion calculation based on the parallel plate model:

(Use PEI: LiClO4 electrolyte gated GFET at the sweep rate of 0.99V/s as an example)

$$n = \frac{\Delta V_{dp} C}{2e} = 0.56V * 1.19 * \frac{\frac{10^{-5}F}{cm^2}}{2e} = 3.33 * 10^{-6} cm^{-2} \qquad equ. 11$$

## 6.3 SEM Microscopic pictures of the back gated GFET:

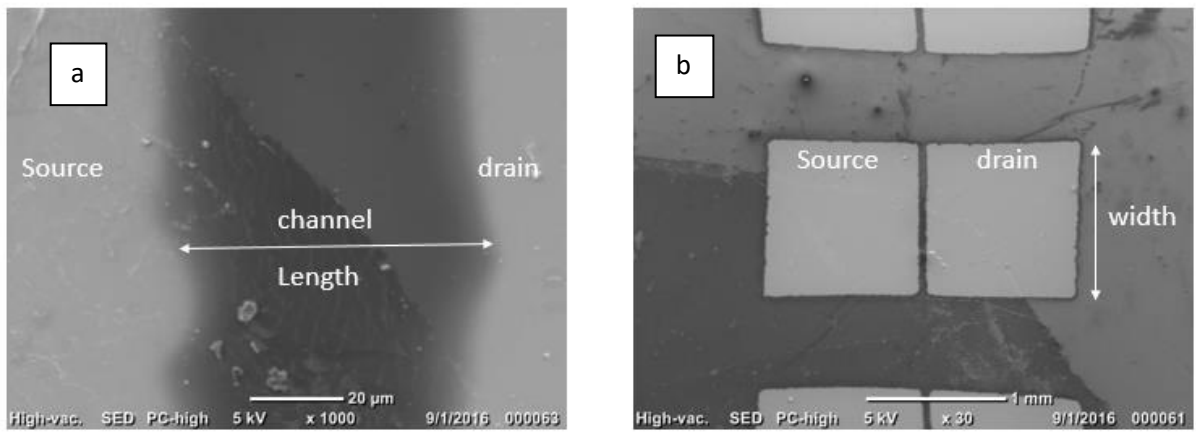

*Figure 5: SEM pictures of the back gated GFET device at 1000x magnification(a); at 30x magnification(b)*

## 6.4 the Field Effect Mobility of back gated device calculation:

The field effect mobility:

$$\mu_{fe} = \frac{dI}{dV_g} \frac{L}{W V_d C_G} \qquad equ. 12$$

The field effect mobility vs. the gate voltage was plotted as follows for the intrinsic graphene back gated GFET

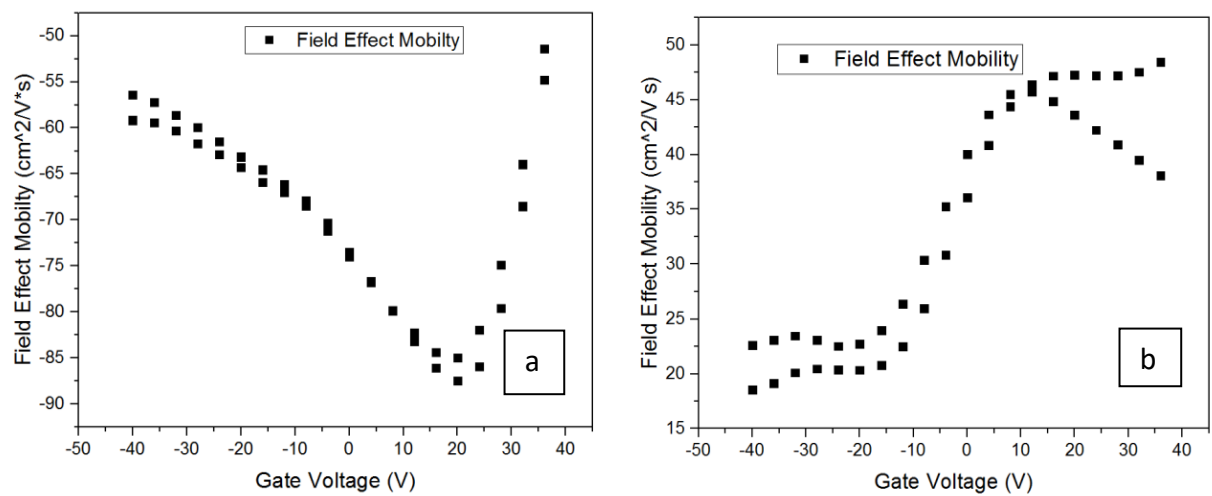

*Figure 6: (a) Field Effect Mobility vs. Vg plot of un-doped back gated Graphene GFET at the sweep rate of 29.98V/s. (b) Field Effect Mobility vs. Vg plot of PEI-doped back gated Graphene GFET at the sweep rate of 28.97V/s. This is extremely lower than the literature values of 6000-10000 cm^2/V\*s [16]*

.